\documentclass[12pt]{article}

\author{Yu.~M.~Zinoviev
       \thanks{E-mail address: yurii.zinoviev@ihep.ru} \\
        {\it Institute for High Energy Physics} \\
        {\it Protvino, Moscow Region, 142280, Russia}}
\title{On Dual Formulation of Gravity. II.\\
	Metric and Affine Connection.}

\date{}

\begin{document}

\maketitle

\begin{abstract}
In this note we construct a dual formulation of gravity where the
main dynamical object is affine connection. We start with the well
known first order Palatini formulation but in (Anti) de Sitter 
space instead of flat Minkowski space as a background. The final
result obtained by solving equations for the metric is the Lagrangian
written by Eddington in his book in 1924. Also there is an
interesting connection with  attempts to construct gravitational
analog of Born-Infeld electrodynamics.
\end{abstract}

\thispagestyle{empty}
\newpage
\setcounter{page}{1}

In general, by dual formulation we mean any situation where the very
same particle is described by different tensor fields. The most
simple and straightforward way to obtain such dual formulation based
on the use of first order "parent" Lagrangians. As is well known in
flat Minkowski space such dualization procedure leads to different
results for massive and massless particles. At the same time in (Anti)
de Sitter space-time gauge invariance requires introduction quadratic
mass-like terms into the Lagrangians. As a result dualization for
massless particles in (Anti) de Sitter spaces \cite{MV04} goes exactly
in the same way as that for massive particles. As an example, we have
recently shown \cite{Zin05a} that by using well known tetrad
formalism it is possible to obtain dual formulation of gravity with
the Lorentz connection being the main dynamical field while tetrad is
just auxiliary fields which could be expressed in terms of Lorentz
connection and its derivatives. But there exist another well known
first order formalism for gravity usually called Palatini formalism,
the main components being the metric and affine connection. Such
formalism differs drastically from the tetrad one because affine
connection is not a gauge invariant object (or, geometrically, it is
not a covariant tensor) and does not have its own gauge invariance.
In spite of this difference, as we are going to show in this note, it
is also possible to apply the same dualization procedure to obtain a
formulation of gravity where the main dynamical field is the affine
connection. Rather naturally and at the same time surprisingly the
final result is nothing else but the Lagrangian written by Eddington
in 1924 \cite{Ed1924}!

Let us start with the first order Lagrangian describing free massless
spin-2 particle in flat Minkowski space:
\begin{equation}
{\cal L}_0 = h^{\mu\nu} ( \partial_\alpha \Gamma_{\mu\nu}{}^\alpha -
\partial_\mu \Gamma_\nu ) + \eta^{\mu\nu} ( \Gamma_{\mu\nu}{}^\alpha
\Gamma_\alpha - \Gamma_{\mu\alpha}{}^\beta \Gamma_{\nu\beta}{}^\alpha)
\end{equation}
Here $h_{\mu\nu}$ is symmetric second rank tensor while
$\Gamma_{\mu\nu}{}^\alpha$ is assumed to be symmetric on the lower
pair of indices. We denote $\Gamma_\alpha =
\Gamma_{\alpha\beta}{}^\beta$, $\Gamma^\alpha = \eta^{\mu\nu}
\Gamma_{\mu\nu}{}^\alpha$ (note, that $\Gamma_\alpha$ and
$\Gamma^\alpha$ are in general different objects). This Lagrangian is
invariant under the following local gauge transformations:
\begin{equation}
\delta h_{\mu\nu} = \partial_\mu \xi_\nu + \partial_\nu \xi_\mu -
\eta_{\mu\nu} (\partial \xi), \qquad \delta \Gamma_{\mu\nu}{}^\alpha
= - \partial_\mu \partial_\nu \xi^\alpha
\end{equation}
As is well known, if one solves the algebraic equation of motion for
the $\Gamma$ field and put the result back into the Lagrangian one
obtains usual second order Lagrangian for the symmetric tensor
$h_{\mu\nu}$. In order to have a possibility to construct dual
formulation where the main dynamical object is $\Gamma$ we move from
the flat Minkowski space to (Anti) de Sitter space. Let
$\bar{g}_{\mu\nu}$ be a metric for this space (it is not a dynamical
quantity, just a background field here) and $D_\mu$ --- derivatives
covariant with respect to background connection winch is torsionless
and metric compatible:
\begin{equation}
D_\alpha \bar{g}_{\mu\nu} = 0, \qquad [ D_\mu, D_\nu ] v_\alpha =
\bar{R}_{\mu\nu,\alpha}{}^\beta (\bar{g}) v_\beta = \kappa (
\bar{g}_{\mu\alpha} \delta_\nu{}^\beta - \delta_\mu{}^\beta
\bar{g}_{\nu\alpha} ) v_\beta
\end{equation}
where $\kappa = - 2 \Lambda/(d-1)(d-2)$.
First of all we have to replace in the Lagrangian as well as in the
gauge transformations the flat metric $\eta_{\mu\nu}$ by
$\bar{g}_{\mu\nu}$ and partial derivatives $\partial_\mu$ by covariant
ones $D_\mu$:
\begin{eqnarray}
{\cal L}_0 &=& h^{\mu\nu} ( D_\alpha \Gamma_{\mu\nu}{}^\alpha -
D_\mu \Gamma_\nu ) + \bar{g}^{\mu\nu} ( \Gamma_{\mu\nu}{}^\alpha
\Gamma_\alpha - \Gamma_{\mu\alpha}{}^\beta
\Gamma_{\nu\beta}{}^\alpha) \nonumber \\
\delta h_{\mu\nu} &=& D_\mu \xi_\nu + D_\nu \xi_\mu -
\bar{g}_{\mu\nu} (D \xi), \qquad \delta \Gamma_{\mu\nu}{}^\alpha
= - \frac{1}{2} ( D_\mu D_\nu + D_\nu D_\mu ) \xi^\alpha
\end{eqnarray}
Now (just because covariant derivatives do not commute) our
Lagrangian is not invariant under the gauge transformations. Indeed,
simple calculations give:
$$
\delta {\cal L}_0 = \kappa [ (d-2) \Gamma_\mu \xi^\mu - \frac{d-3}{2}
\Gamma^\mu \xi_\mu - \frac{3d-1}{2} h^{\mu\nu} D_\mu \xi_\nu + h
(D \xi) ]
$$
But gauge invariance could be easily restored by adding terms
quadratic in $h_{\mu\nu}$ field to the Lagrangian as well as
appropriate corrections for the gauge transformations:
\begin{eqnarray}
\Delta {\cal L}_0 &=& \frac{\kappa (d-1)}{2} [ h^{\mu\nu} h_{\mu\nu}
- \frac{1}{d-2} h^2 ] \nonumber \\
\delta' \Gamma_{\mu\nu}{}^\alpha &=& \frac{\kappa}{2}
(\delta_\mu{}^\alpha \xi_\nu + \delta_\nu{}^\alpha \xi_\mu ) - \kappa
\bar{g}_{\mu\nu} \xi^\alpha
\end{eqnarray}
Now one can easily solve the equations for the $h_{\mu\nu}$ field,
which are also algebraic now, to obtain:
\begin{equation}
h_{\mu\nu} = \frac{1}{\kappa(d-1)} [ R_{\mu\nu} - \frac{1}{2}
\bar{g}_{\mu\nu} R ]
\end{equation}
where we introduced a symmetric second rank tensor (it is not a full
Ricci tensor yet, only the first part of it):
\begin{equation}
R_{(\mu\nu)} = \frac{1}{2} ( D_\mu \Gamma_\nu + D_\nu \Gamma_\mu) -
D_\alpha \Gamma_{\mu\nu}{}^\alpha
\end{equation}
Then if we put this expression back into the initial first order
Lagrangian we obtain dual second order formulation for massless
spin-2 particle in terms of $\Gamma$ field:
\begin{equation}
{\cal L}_{II} = - \frac{1}{\kappa(d-1)} [ R^{\mu\nu} R_{\mu\nu} -
\frac{1}{2} R^2 ] + \bar{g}^{\mu\nu} ( \Gamma_{\mu\nu}{}^\alpha
\Gamma_\alpha - \Gamma_{\mu\alpha}{}^\beta \Gamma_{\nu\beta}{}^\alpha)
\end{equation}
A natural question arises: our field $\Gamma_{\mu\nu}{}^\alpha$ has a
lot of independent components (40 in $d=4$ instead of two helicities
for massless spin-2 particle), so there should exist a large gauge
symmetry in such a model. And indeed, it is easy to check that the
kinetic terms in our second order Lagrangian are invariant under the
local "affine" transformations:
\begin{eqnarray}
\delta \Gamma_{\mu\nu}{}^\alpha &=& \partial_\mu z_\nu{}^\alpha +
\partial_\nu z_\mu{}^\alpha + \frac{1}{d-1} [ \delta_\mu{}^\alpha
(\partial z)_\nu + \delta_\nu{}^\alpha (\partial z)_\mu ] - \nonumber
\\
 && - \frac{1}{d-1} [ \delta_\mu{}^\alpha \partial_\nu z +
\delta_\nu{}^\alpha \partial_\mu z ]
\end{eqnarray}
where $z_\mu{}^\nu$ is arbitrary second rank tensor and $z =
z_\mu{}^\mu$.

Now, having in our disposal an alternative description for massless
spin-2 particle, it is natural to see how an interaction in such
dual theory looks like. Nice feature of Palatini formulation is that
switching on an interaction is a simple one step procedure
\cite{Des70}. But as we have seen, it is very important for the
possibility to construct dual formulations to work not in a flat
Minkowski space but in (Anti) de Sitter space. So we start with the
usual Lagrangian with the cosmological term:
\begin{equation}
{\cal L} = \sqrt{- g} g^{\mu\nu} R_{\mu\nu} + \Lambda \sqrt{- g}
\end{equation}
where now
\begin{equation}
R_{\mu\nu} = \frac{1}{2} ( D_\mu \Gamma_\nu + D_\nu \Gamma_\mu) -
D_\alpha \Gamma_{\mu\nu}{}^\alpha + \Gamma_{\mu\nu}{}^\alpha
\Gamma_\alpha - \Gamma_{\mu\alpha}{}^\beta \Gamma_{\nu\beta}{}^\alpha
\end{equation}
Then we introduce a convenient combination $\hat{g}^{\mu\nu} =
\sqrt{-g} g^{\mu\nu}$ and rewrite a Lagrangian as:
\begin{equation}
{\cal L} = \hat{g}^{\nu\nu} R_{\mu\nu} + \Lambda
det(\hat{g}^{\mu\nu})^{\frac{1}{d-2}}
\end{equation}
The crucial point here is that first term contains $\hat{g}$ only
linearly. As a result it is possible to get complete nonlinear
solution of the $\hat{g}$ equations. We obtain (up to some numerical
coefficients):
\begin{equation}
\hat{g}^{\mu\nu} \simeq \sqrt{det(R_{\mu\nu})} (R^{\mu\nu})^{-1}
\end{equation}
At last, if we put this expression back into the first order
Lagrangian we obtain (again up to normalization) a very simple and
elegant Lagrangian:
\begin{equation}
{\cal L} = \sqrt{det(R_{\mu\nu})}
\end{equation}
And it is just a Lagrangian written by Eddington eighty years ago in
his book \cite{Ed1924}! This result is very natural because this
Lagrangian is the only invariant that could be constructed out of the
affine connection alone, without any use of metric or any other
objects, but it is exiting that this Lagrangian turns out to be dual
formulation of usual gravity theory. Let us stress once again that
working in a flat Minkowski space it is very hard if at all possible
to give any reasonable physical interpretation to such model. But let
us consider this model on a (Anti) de Sitter background. For that
purpose we represent a total connection as
$\Gamma_{\mu\nu}{}^\alpha = \bar{\Gamma}_{\mu\nu}{}^\alpha +
\tilde{\Gamma}_{\mu\nu}{}^\alpha $
where $\bar{\Gamma}_{\mu\nu}{}^\alpha$ is a background connection
while $\tilde{\Gamma}_{\mu\nu}{}^\alpha$ --- small perturbation
around it (see e.g. \cite{BG99,Pet05}). Then for the curvature tensor
we will have:
\begin{equation}
R_{\mu\nu,\alpha}{}^\beta = \bar{R}_{\mu\nu,\alpha}{}^\beta + [ D_\mu
\tilde{\Gamma}_{\nu\alpha}{}^\beta +
\tilde{\Gamma}_{\mu\alpha}{}^\rho \tilde{\Gamma}_{\rho\nu}{}^\beta -
(\mu \leftrightarrow \nu) ]
\end{equation}
where $\bar{R}_{\mu\nu,\alpha}{}^\beta$ is a curvature tensor for the
background connection and $D_\mu$ is a derivative covariant with
respect to $\bar{\Gamma}$. Then for the constant curvature space we
have $\bar{R}_{\mu\nu} = \Lambda \bar{g}_{\mu\nu}$ so the Lagrangian
takes the form:
\begin{equation}
{\cal L} = \sqrt{det(\Lambda \bar{g}_{\mu\nu} + \tilde{R}_{\mu\nu})}
\end{equation}
It is interesting that Lagrangians of such kind have already been
investigated e.g. \cite{DG98,Vol03,Woh03,Nie04,Com05,Vol05} in
attempts to construct gravitational analog of the Born-Infeld
electrodynamics. But now the interpretation of the Lagrangian is
drastically different. Indeed, let us use the well known decomposition
for the determinant
$$
\sqrt{det(I + A)} = 1 + \frac{1}{2} Sp(A) + \frac{1}{8} (Sp(A))^2 -
\frac{1}{4} Sp(A^2) + ....
$$
where $A$ is any matrix. Then if we consider the curvature
$R_{\mu\nu}$ as being expressed in terms of metric and its second
derivatives the first linear terms gives scalar curvature while
quadratic terms give higher derivative terms leading to the
appearance of ghosts. But here the main dynamical quantity is affine
connection $ \Gamma$ and the curvature $R_{\mu\nu}$ contains only
first derivatives. As a result a term linear in $R$ is just a total
derivative and could be dropped out of the action, while the
quadratic terms give exactly the kinetic terms we obtained above.

Finally, let us add some comments on possible interaction with matter
in such formulation of gravity. The most clear and straightforward way
to obtain these interactions is to start with usual interactions in
first order form and then try to go to the dual formulation. For
example, for the scalar field we get:
\begin{eqnarray}
{\cal L} &=& \sqrt{-g} [ g^{\mu\nu} R_{\mu\nu} + \Lambda +
\frac{1}{2} g^{\mu\nu} \partial_\mu \varphi \partial_\nu \varphi -
\frac{m^2}{2} \varphi^2 ] = \nonumber \\
 &=& \hat{g}^{\mu\nu} ( R_{\mu\nu} + \frac{1}{2} \partial_\mu \varphi
\partial_\nu \varphi ) + ( \Lambda - \frac{m^2}{2} \varphi^2 ) 
det(\hat{g}^{\mu\nu})^{\frac{1}{d-2}}
\end{eqnarray}
and the second line shows that the main effect is the replacement of
$R_{\mu\nu}$ by $R_{\mu\nu} + \frac{1}{2} \partial_\mu \varphi
\partial_\nu \varphi$ (compare \cite{Vol05}). Also if scalar field
has nonzero mass the cosmological constant $\Lambda$ is replaced by
field dependent combination $\Lambda - \frac{m^2}{2} \varphi^2$.
But for the vector field (even massless) the situation turns out to
be much more complicated because even for the minimal interaction:
\begin{eqnarray}
{\cal L} &=& \sqrt{-g} [ g^{\mu\nu} R_{\mu\nu} + \Lambda - \frac{1}{4}
g^{\mu\alpha} g^{\nu\beta} F_{\mu\nu} F_{\alpha\beta} ] = \nonumber \\
 &=& \hat{g}^{\nu\nu} R_{\mu\nu} + \Lambda
det(\hat{g}^{\mu\nu})^{\frac{1}{d-2}} - \frac{1}{4}
det(\hat{g}^{\mu\nu})^{-\frac{1}{d-2}} \hat{g}^{\mu\alpha}
\hat{g}^{\nu\alpha} F_{\mu\nu} F_{\alpha\beta}
\end{eqnarray}
equations for the $\hat{g}$ become highly nonlinear. But in a weak
field approximation such model could reproduce a correct kinetic term
for the vector field. Note also that the corrections to the
$R_{\mu\nu}$ tensor here start with the terms quadratic in
$F_{\mu\nu}$ and there is no term linear in it in contrast with
\cite{Vol05}.

Thus we have shown that the dualization procedure based on the use of
(Anti) de Sitter background space could be applied to the gravity
theory in a Palatini formalism and leads to the formulation in terms
of affine connection. In this, the final Lagrangian coincides with
that of Eddington \cite{Ed1924}. A number of interesting question
arises, for example, whose related with the gauge symmetries of such
formulation, which deserve further study.

\end{document}